\documentclass{aa}  

\usepackage{multicol}
\usepackage{graphicx}
\usepackage{multirow}
\usepackage{amsmath}
\usepackage{txfonts}
\usepackage{natbib}
\bibpunct{(}{)}{;}{a}{}{,} 
\def\ltsim{\raise 2pt \hbox {$<$} \kern-1.1em \lower 4pt \hbox {$\sim$}}
\def\gtsim{\raise 2pt \hbox {$>$} \kern-1.1em \lower 4pt \hbox {$\sim$}}

\begin{document} 

\title{
The great Kite in the sky:\\a LOFAR observation of the radio source in Abell~2626}

\author{A. Ignesti\inst{1,2}\thanks{Contact: {\tt alessandro.ignesti2@unibo.it}}, 
T. Shimwell\inst{3,4}, 
G. Brunetti\inst{2}, 
M. Gitti\inst{1,2}, 
H. Intema\inst{4,6}, 
R. J. van Weeren\inst{4}, 
M. J. Hardcastle\inst{5},  
A. O. Clarke\inst{7,8}, 
A. Botteon\inst{4}, 
G. Di Gennaro\inst{4}, 
M. Br\"uggen\inst{9}, 
I. Browne\inst{8}, 
S. Mandal\inst{4}, 
H. J. A. R\"ottgering\inst{4}, 
V. Cuciti\inst{9}, 
F. de Gasperin\inst{9},
R. Cassano\inst{2}, 
A. M. M. Scaife\inst{8}}

\institute{
  Dipartimento di Fisica e Astronomia, Universit\`a di Bologna, via Gobetti 93/2, 40129 Bologna, Italy
\and 
INAF, Istituto di Radioastronomia di Bologna, via Gobetti 101, 40129 Bologna, Italy
\and
ASTRON, the Netherlands Institute for Radio Astronomy, Postbus 2, 7990 AA Dwingeloo, The Netherlands
\and
Leiden Observatory, Leiden University, PO Box 9513, 2300 RA Leiden, The Netherlands
\and
Centre for Astrophysics Research, University of Hertfordshire, College Lane, Hatfield AL10 9AB, UK
\and
International Centre for Radio Astronomy Research, Curtin University, GPO Box U1987, Perth,WA 6845, Australia
\and
SKA Organisation, Jodrell Bank Observatory, SK11 9DL, UK
\and
Jodrell Bank Centre for Astrophysics, Department of Physics $\&$ Astronomy, University of Manchester, Oxford Road, Manchester M13 9PL, UK
\and
Hamburger Sternwarte, Universit\"at Hamburg, Gojenbergsweg 112, D-21029, Hamburg, Germany }
\authorrunning{Ignesti et al.}
\titlerunning{The great Kite in the sky: a LOFAR observation of the radio source in Abell~2626}

\date{Accepted on 23/09/2020}

\abstract 
{
 The radio source at the center of the galaxy cluster Abell 2626, also known as the Kite, stands out for its unique morphology composed of four, symmetric arcs. Previous studies have probed the properties of this source at different frequencies and its interplay with the surrounding thermal plasma, but the puzzle of its origin is still unsolved.}
{
We use new LOw Frequency ARray (LOFAR) observation from the LOFAR Two-meter Sky Survey at 144 MHz to investigate the origin of the Kite.}
{
  We present a detailed analysis of the new radio data which we combined with archival radio and X-ray observations. We have produced a new, resolved spectral index map of the source with a resolution of 7$''$ and we studied the spatial correlation of radio and X-ray emission to investigate the interplay between thermal and non-thermal plasma.}
{
  The new LOFAR data have changed our view of the Kite by discovering two steep-spectrum ($\alpha<-1.5$) plumes of emission connected to the arcs. The spectral analysis shows, for the first time, a spatial trend of the spectrum along the arcs with evidence of curved synchrotron spectra and a spatial correlation with the X-ray surface brightness. On the basis of our results, we propose that the Kite was originally an X-shaped radio galaxy whose fossil radio plasma, after the end of the activity of the central active galactic nucleus, has been compressed due to motions of the thermal plasma in which it is encompassed. The interplay between the compression and advection of the fossil plasma, with the restarting of the nuclear activity of the central galaxy, could have enhanced the radio emission of the fossil plasma producing the arcs of the Kite. We present also the first, low-frequency observation of a jellyfish galaxy in the same field, in which we detect extended, low-frequency emission without a counterpart at higher frequencies.
 } {}


\maketitle
\section{Introduction}
Amongst the radio sources observed in galaxy clusters, the one at the center of the relaxed galaxy cluster Abell 2626 (hereafter A2626) stands out for its unique properties. The discovery of diffuse radio emission at the center of the cluster surrounding the radio source 3C464 was reported for the first time by \citet[][]{Rizza_2000}. In the first image at 1.4 GHz the source resembled an amorphous 100 kpc blob around the central galaxy IC5338. Thus it was classified as a radio mini-halo \citep[][]{Gitti_2004}, a class of diffuse radio sources observed at the centers of relaxed clusters. They also noted the presence of two symmetric substructures embedded in the diffuse emission around the central galaxy. On the basis of a detailed analysis of the X-ray emission and the presence of two optical nuclei at the center of the galaxy, \citet[][]{Wong_2008} suggested that these symmetric features could be the remnants of past activity of the central active galactic nucleus (AGN), where the precession of the AGN might have left two fossil plasma trails. Later, deeper high-resolution observations carried out with the Very Large Array (VLA) in A+B configuration at 1.4 GHz found that the source was not amorphous, but instead most of the emission comes from three elongated and collimated structures that were called $``$arcs" \citep[][]{Gitti_2013b}. The arcs were observed in the northward (N), westward (W) and southward (S) of the central galaxy, with a junction between the southern and western ones (Figure \ref{canvas}, top-right panel). The most puzzling features were the remarkably symmetric northern and southern arcs and their concavity, which was directed outward and earned this source the name of the $``$Kite". In these 1.4 GHz images the major and minor axis are 120 and 40 kpc, respectively. The minimum, projected distance of the arcs from the central AGN is $\sim25$ kpc. The discovery of the third arc suggested a more complex scenario than the one invoked by \citet[][]{Wong_2008} and the properties that were emerging discouraged the classification of the Kite as one of the common diffuse radio sources observed in galaxy clusters, such as radio relics or radio mini-halos. New insights into its origin came from further radio observations at a different frequency. By analyzing archival Giant Metrewave Radio Telescope (GMRT) data at 610 MHz, \citet[][]{Kale_2017} discovered the presence of a fainter fourth arc on the east (E), thus completing the symmetry of the system (Figure \ref{canvas}, top-left panel). They produced a spectral index map between 610 MHz and 1.4 GHz, measuring a spectral index in the arcs of $\alpha\sim -2.5$ (we define the radio spectrum $S\propto\nu^{\alpha}$ where $S$ is the observed radio flux density at frequency $\nu$). Subsequently, \citet[][]{Ignesti_2017} presented Jansky Very Large Array images of the Kite obtained at 3.0 and 5.5 GHz in C-configuration. They found that the arcs have a steep spectral index $\alpha\sim-3$ up to 3 GHz, with no clear evidences of spectral index gradient along the arc length. These results severely constrained the jet-precession scenario because if the arcs were created gradually by the AGN, then a trend in spectral index was expected due the radiative ageing of the plasma. \citet[][]{Ignesti_2018} carried out a detailed study of the deepest {\it Chandra} observation available of the cluster. They discovered a cold front at the center of the cluster and they found that the edge of the sloshing coincides remarkably well with the junction between the S and W arcs. Subsequent studies confirmed the presence of a cavity in the intra-cluster medium (ICM) located beyond the S-W junction of the radio arcs and suggested that it was created by past AGN activity \citep[][]{Kadam_2019}.\\

In this paper we present the results of the analysis of a new observation of this cluster made with the LOw-Frequency ARay (LOFAR). This observation is part of the LOFAR Two-metre Sky Survey \citep[LoTSS][]{Shimwell_2017}. This survey aims to observe the entire northern sky at 120-168 MHz with a resolution of $\sim6''$ and a sensitivitiy of $\sim100$ $\mu$Jy beam$^{-1}$. In addition to the new LOFAR observations, we also combine with archival radio (GMRT and VLA) and X-ray ({\it Chandra} and {\it XMM-Newton}) to get new insights on the nature of the Kite. The paper is structured as follows. In Section \ref{data_prep} we present the processing of the new LOFAR data and the archival radio and X-ray observations. The results are analyzed in Section \ref{anl}, in which we also present an analysis of the low-frequency emission of the jellyfish galaxy JW100, and discussed in Section \ref{disc}. \\

We adopt a $\mathrm{\Lambda CDM}$ cosmology with $\mathrm{H_{0}=70}$ km $\mathrm{s^{-1}Mpc^{-1}}$, $\Omega_\text{M} = 1 - \Omega_{\Lambda} = 0.3$. At the cluster redshift \citep[$z$=0.0553, ][]{Struble-Rood_1999}, it yields a luminosity distance of 246.8 Mpc and 1 arcsec = 1.1 kpc\footnote{{\ttfamily
http://www.astro.ucla.edu/$\#$7Ewright/CosmoCalc.html}}.
 \section{Data preparation}
 \label{data_prep}
\subsection{Radio data}
\begin{figure*}
\centering
\includegraphics[width=.95\textwidth]{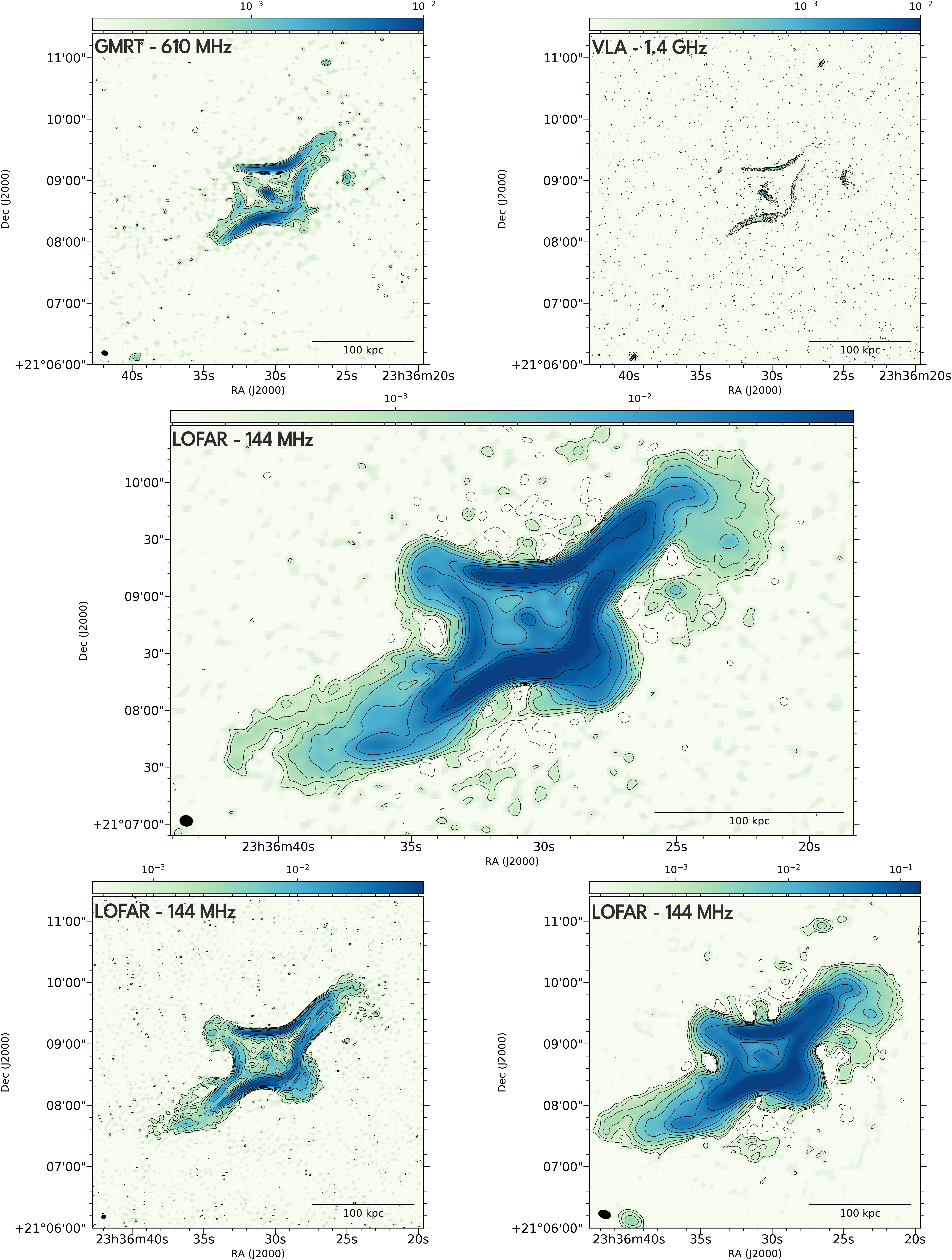}
\caption{\label{canvas}Images of the Kite at different frequencies obtained by re-processing archival datasets. Top-left: GMRT image at 610 MHz, the RMS is 120 $\mu$Jy beam$^{-1}$ and the resolution is 6.2$''\times$4.1$''$; Top-right: VLA A+B image at 1.4 GHz, the RMS is 13 $\mu$Jy beam$^{-1}$ and the resolution is 1.5$''\times$1.4$''$. In both image the contours are at the -3, 3, 6, 12 and 24 $\sigma$ levels. New images at 144 MHz. Middle: {\ttfamily robust=-0.75}, the final resolution is 6.7$''$$\times$5.6$''$ and the RMS is 120 $\mu$Jy beam$^{-1}$. The contours are at the -3, 3, 6, 12, 24, 48, 96, 192, 384 $\sigma$ level; Bottom-left: {\ttfamily robust=-2}, the final resolution is 4.2$''$$\times$3.2$''$ and the RMS is 380 $\mu$Jy beam$^{-1}$. The contours are at the -3, 3, 6, 12, 24, 48, 96 $\sigma$ level; Bottom-right: {\ttfamily robust=-0.25}, the final resolution is 12.2$''\times$7.7$''$ and the RMS is 140 $\mu$Jy beam$^{-1}$. The contours are at the -3, 3, 6, 12, 24, 48, 96, 192, 384 $\sigma$ level. The surface brightness is reported in units of Jy beam$^{-1}$.}
\end{figure*}

In this work we present the analysis of the pointing P353+21 of the LoTSS survey \citep[][]{Shimwell_2017,Shimwell_2019}. The observation was made using  the Dutch High Band Antenna (HBA) array, which operates in the 120-168 MHz band, for a total observation time of 8.33 hrs. We reduced the dataset using the direction-dependent data-reduction pipeline {\scshape ddf-pipeline} v. 2.2 developed by the LOFAR Surveys Key Science Project\footnote{{\tt https://github.com/mhardcastle/ddf-pipeline}}. The data processing makes use of {\scshape prefactor}  \citep[][]{vanWeeren_2016,Williams_2016,deGasperin_2019}, {\scshape killMS} \citep[][]{Tasse_2014,Smirnov_2015} and {\scshape DDFacet} \citep[][]{Tasse_2018} to perform the calibration and imaging of the entire LOFAR field of view. Then, we performed additional phase and amplitude self-calibration cycles to correct the residual artifacts in a smaller region extracted from the observation and centered on the source, where the direction-dependent errors are assumed to be negligible. The details of this extraction and re-calibration step will be discussed in a forth-coming paper (van Weeren et al., in prep.).
We produced images at different resolutions using {\scshape WSClean} v2.6 \citep[][]{Offringa_2014} and using several different Briggs weightings \citep[][]{Briggs_1994}, with {\ttfamily robust} going from 0 to -2, and multi-scale cleaning \citep[][]{Offringa_2017}. An inner uvcut of 80$\lambda$, corresponding to an angular scale of 43$'$, was applied to the data to drop the shortest spacings where calibration is more challenging. The resulting images are shown in Figure \ref{canvas}. Finally, we corrected for the systematic offset of the LOFAR flux density scale produced by inaccuracies in the LOFAR HBA beam model by multiplying it with a factor of 1.17 which aligns it with the flux-scale of the LoTSS-DR2 catalog that is calibrated off the NRAO VLA Sky Survey (NVSS) flux-scale. Following LoTSS \citep[][]{Shimwell_2019}, we adopt a conservative calibration error of 20$\%$, which dominates the uncertainties on the LOFAR flux densities.\\

We present here also the GMRT and the VLA images of A2626 obtained from archival observations. We retrieved the archival GMRT observation of the cluster at 610 MHz, with a time on target of 95 mins, (observation 561, PI Clarke) and we processed it using the {\scshape SPAM} pipeline \citep[see][for details]{Intema_2009,Intema_2017}. For the VLA observation we used the calibrated dataset presented by \citet[][]{Gitti_2013b}. The images at 610 MHz and 1.4 GHz presented in this work were made using {\scshape WSClean}. \\
\subsection{X-ray data}

In this work we analyzed archival {\it Chandra} and {\it XMM-Newton} observations of A2626 to produce images in the 0.5-2.0 keV band of the cluster. Specifically, we used the {\it Chandra} observations 3192 and 16136, for a total exposure time of 130 ks, and the {\it XMM-Newton} observations 0083150201 and 0148310101, for a total exposure time of  55  ks.  We reprocessed the {\it Chandra} datasets with CIAO 4.10 and CALDB 4.8.1 to correct for known time-dependent gain and for charge transfer inefficiency. In order to filter out strong background flares, we also applied screening of the event files\footnote{{\tt http://cxc.harvard.edu/ciao/guides/acis\_data.html}}. For the background subtraction, we used the CALDB $``$Blank-Sky" files normalized to the count rate of the source image in the 10-12 keV band. The data reduction of the XMM-Newton observations was performed using the Extended Source Analysis Software (ESAS) integrated in the Scientific Analysis System (SAS v16.1.0). Observation periods affected by soft proton flares were filtered out with the the mos-filter and pn-filter tasks. For each detector and observation we produced count images, background images, and exposure maps. These were combined to create a background-subtracted and exposure-corrected EPIC mosaic in the 0.5-2.0 keV band. 

\section{Image analysis and Results}
\label{anl}
\begin{figure}
\centering
\includegraphics[width=.5\textwidth]{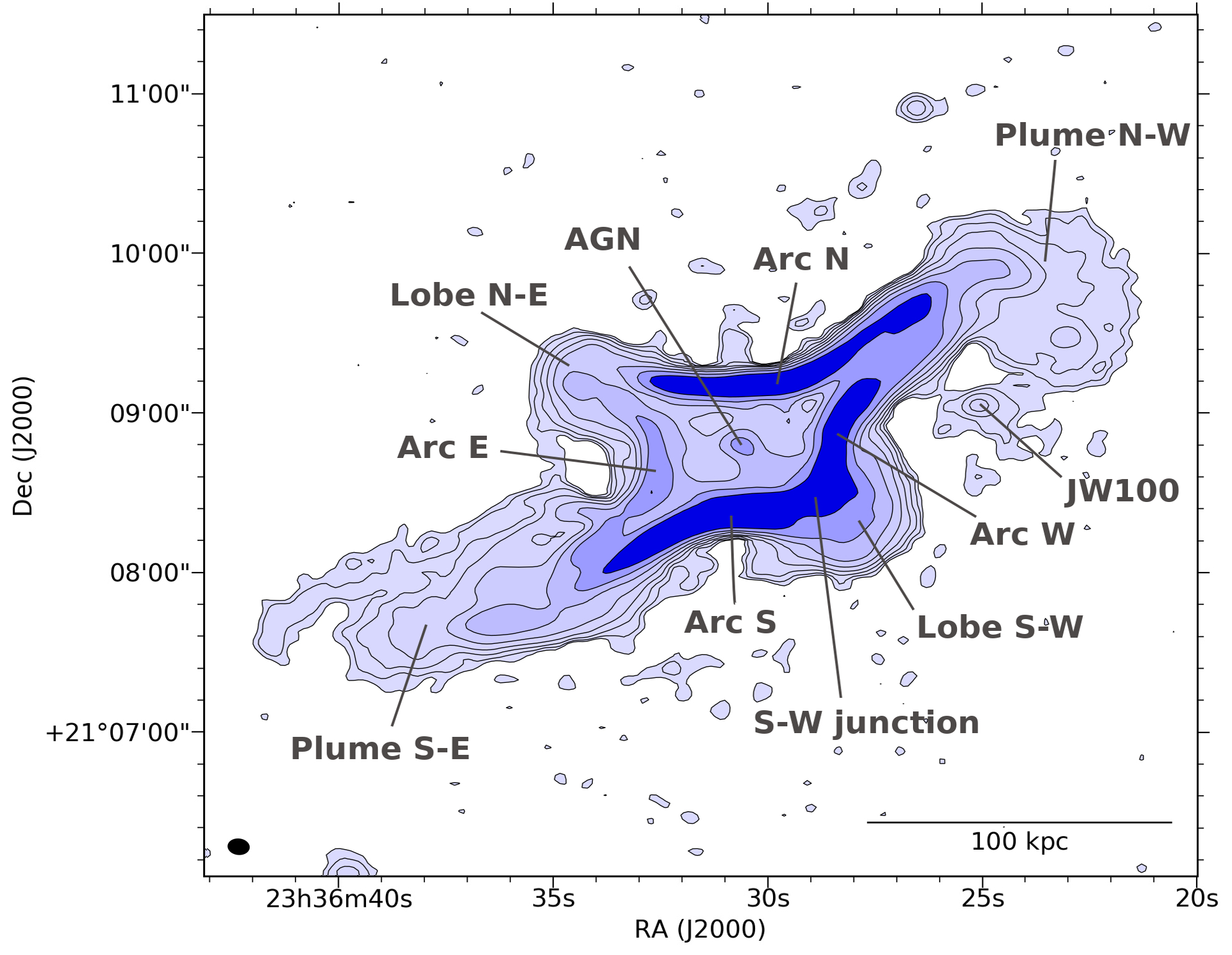}
\caption{\label{scheme} Sketch of the Kite observed at 144 MHz (Figure \ref{canvas}, middle panel) color-coded for the surface brightness level where we indicate the main components of the radio emission.  }
\end{figure}

At 144 MHz the source shows its striking, kite-like morphology regardless of the resolution of the images (Figure \ref{canvas}). The arcs are more extended along the major and minor axis than at higher frequencies, but their inner edges still have a projected distance of $\sim25$ kpc from the AGN and we detect extended emission between the AGN and the arcs. The LOFAR data reveal for the first time two large $``$plumes" of emission emerging from the ends of the arcs in in the south-east and north-west directions (Figure \ref{canvas}, middle and bottom-right panels), more than doubling the projected size of the Kite to $\sim$220 kpc ($\simeq 3.7'$). The new data also reveals two lobes of emission toward N-E and S-W. Finally, we observe extended emission associated with the galaxy JW100 (IC5337, see Section \ref{jw}). In Figure \ref{scheme} is reported a schematic drawing of the Kite where we indicate the several components of the radio emission observed by LOFAR. In the {\ttfamily robust=-0.75} image (Figure \ref{canvas}, middle panel) we measure a 144 MHz flux density of 7.7$\pm$1.6 Jy for the whole source. The surface brightness contours show that the arcs, whose high brightness at low frequeny ($>100\sigma$) was expected due to their steep spectra \citep[][]{Kale_2017, Ignesti_2017}, dominate the total emission of the Kite. \\
\subsection{Spectral index map}
\begin{center}
\begin{figure*}
\centering
\includegraphics[width=\textwidth]{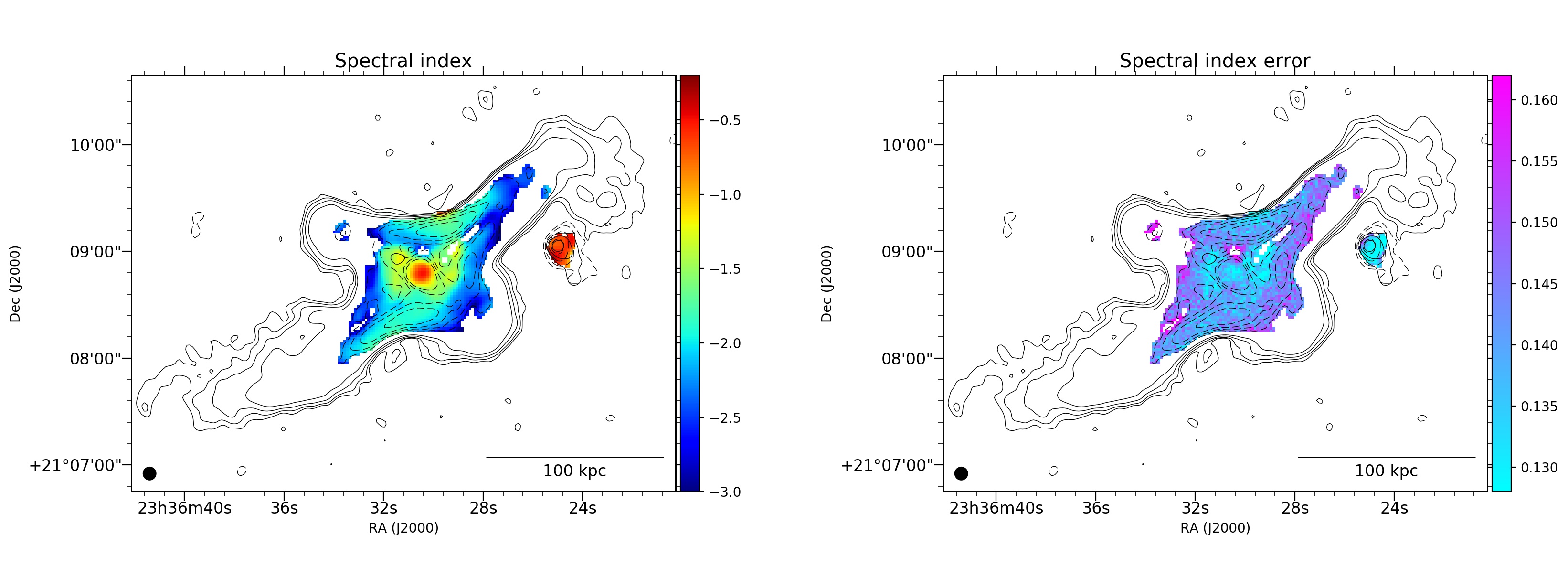}

\caption{\label{canvas_spectral} Spectral index map (left) and relative error map (right) obtained combining LOFAR, GMRT and VLA maps at, respectively, 144, 610 MHz and 1.4 GHz. The resolution is $7''\times7''$. We report the contours at the 3, 6, 12, 24 $\sigma$ levels of the maps used to produce the spectral index map with continuous and dashed lines showing the LOFAR and the VLA image, respectively. }
\end{figure*}
\end{center}

In order to probe the properties of the relativistic electrons, we combined the new data with archival GMRT and VLA observation at 610 MHz and 1.4 GHz, respectively, to produce a resolved, 3-band spectral index map (Fig \ref{canvas_spectral}). First we produced new maps at the three frequencies using {\scshape WSClean} with a {\ttfamily uniform} weighting for the visibilities and a uv-range of 200-46000 $\lambda$  in order to match the same spatial scale in all wavelengths. The lower cut at 200 $\lambda$ implies a largest recoverable angular scale of $\sim20'$, hence it allows the detection of the complete structure of the Kite for each band. The resulting images were smoothed to a common resolution of $7''\times7''$. The resulting RMS noise of the LOFAR, GMRT and VLA maps are, respectively, 280, 150 and 35 $\mu$Jy beam$^{-1}$. We combined the three maps to produce a spectral index map and the corresponding error map. We compared the emission above the 3$\sigma$ level of the LOFAR image with the emission above 2$\sigma$ for the GMRT and VLA images to probe the steep-spectrum emission in the arcs. For each pixel of each map, we measured the flux density and its associated error. For each frequency, we extracted random values of the flux within the associated error by assuming a normal distribution. Then we fitted the three boot-strapped flux densities with a power-law spectrum $S(\nu)=S_0\nu^{\alpha}$ to evaluate the spectral index. We repeated this procedure 500 times for each pixel, thus ending with a normal-like distribution of values of $\alpha$ whose skewness depends on the relative uncertainties of the flux densities in each pixel. The final spectral index map in Figure \ref{canvas_spectral} shows in each pixel the mean of these distributions, whereas the error map shows the standard deviations. This approach allows to get reliable errors on the spectral index that reflect the uncertainties of the three measurements.\\

This spectral index map allows us to divide the radio source in two regions. The central region within the arcs shows a flatter spectrum, $-1<\alpha<-0.5$, whereas in the arcs the spectral index is $\alpha<-1.5$. Thanks to the resolution and sensitivity provided by the new LOFAR data, we observe, for the first time, a spatial gradient along the arcs N and S, where the spectrum steepens moving from the centers of the arcs to their ends. On the contrary, the arcs E and W do not show this trend, instead they exhibit more uniform and steeper spectra. Despite the low threshold in surface brightness at the higher frequencies, we could not map the spectral index of the lobes and the plumes. For the latter, we could derive only an upper limit for the spectral index $\alpha<-$1.5 by comparing their mean surface brightness at 144 MHz (in untis of Jy beam$^{-1}$) with the 2$\sigma$ level of VLA image. Finally we used the LOFAR and VLA images produced for the spectral index analysis to estimate the integrated spectral index of the Kite. By comparing the total flux densities within the 3$\sigma$ contours of the 144 MHz map, we estimated an integrated spectral index $\alpha=-2.0\pm0.4$ between 144 MHz and 1.4 GHz, that entails a k-corrected radio power at 144 MHz of 6$\times 10^{25}$ W Hz$^{-1}$.\\  
\subsection{Point-to-point analysis}
\begin{figure}
\centering
 \includegraphics[width=.5\textwidth]{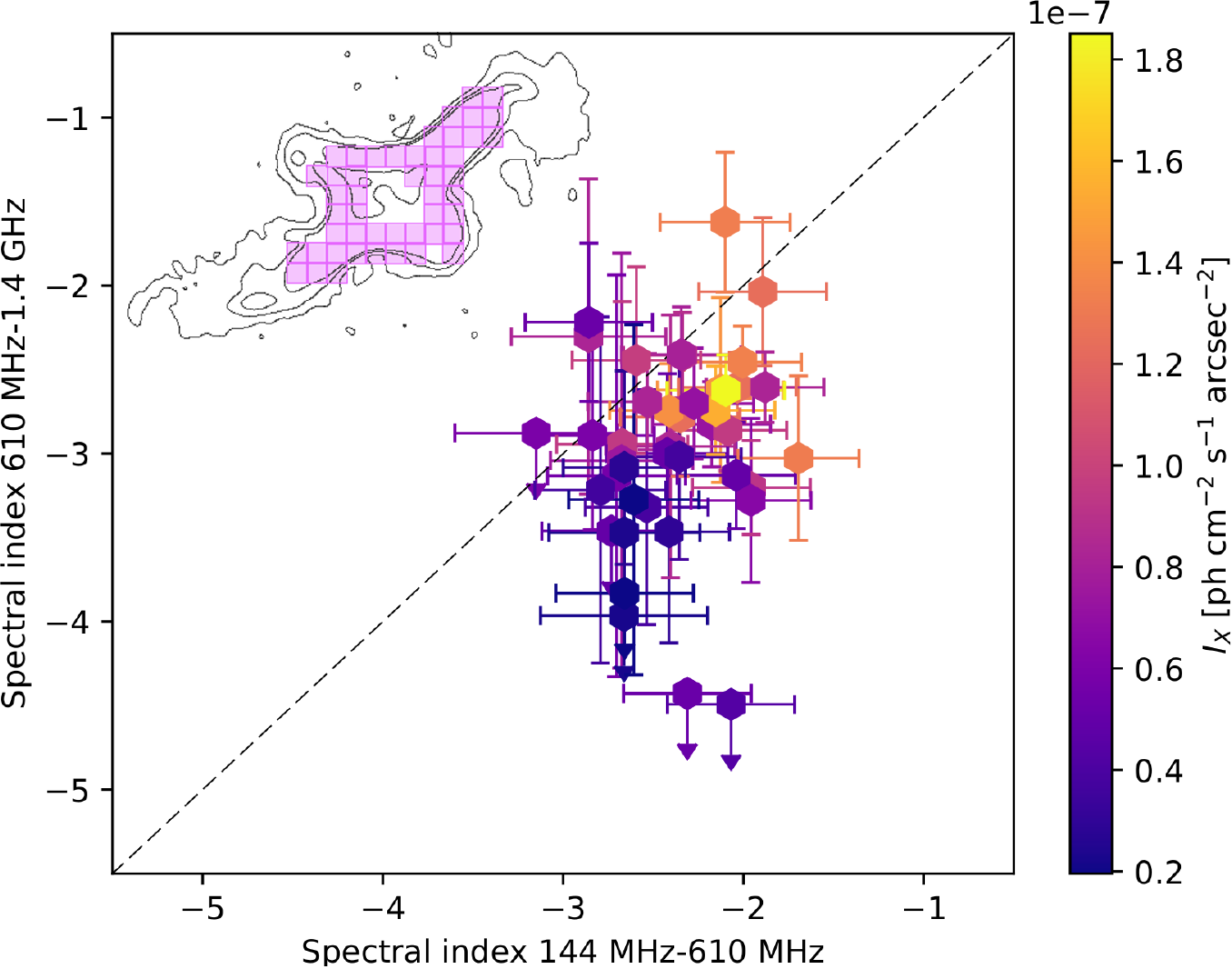}
 \caption{\label{ptp} Spectral index in 610 MHz-1.4 GHz vs spectral index in 144 MHz-610 MHz band, color-coded for the X-ray surface brightness $I_\text{X}$. The values were measured in the cells of the green grid that is shown in the top-left corner, overlapped on the 3, 24, 96$\sigma$ levels of the 144 MHz image involved in the spectral analysis. Cells where the relative error on the spectral index is greater than 50$\%$ are shown as upper limits.}
\end{figure}

We performed a point-to-point analysis to explore the spatial correlation between the spectral index, $\alpha$, and X-ray surface brightness, $I_\text{X}$, which is a proxy of the projected thermal plasma density. Observing (or not) a spatial correlation between the two quantities could potentially provide an insight into the intrinsic relation between the thermal plasma and the non-thermal ICM components, i.e. the magnetic field and the relativistic particles. In this work we combined the observations of the cluster at 144, 610 MHz and 1.4 GHz with the {\it Chandra} X-ray image. We sampled the region of the Kite where the 610 MHz emission is above the 3$\sigma$ level with a grid whose cells are 12$''\times12''$ in size to have a compromise between the signal-to-noise of each cell and the number of significant points. On the one hand, this sampling assures a reliable estimate of the spectral index for the low-band (144-610 MHz) of the spectrum in each cell. On the other hand, it introduces large uncertainties in the high-band (610 MHz-1.4 GHz) for those regions where the radio surface brightness at 1.4 GHz is below $3\sigma$. The grid includes the central region (except the AGN and the jet-like feature which we excluded from this analysis), the arcs and part of the emission in the NW-SE direction (see Figure \ref{ptp}). For each cell of the grid, we measured the flux densities at the 3 frequencies and the $I_\text{X}$ from the {\it Chandra} image of the cluster to produce the color-color plot (Figure \ref{ptp}). We observe that the majority of the points cover the Kite are located below the 1:1 line, which implies that the spectrum of the radio emission is steeper in the 610 MHz-1.4 GHz band than in the 144 MHz - 610 MHz one for the majority of the Kite, i.e. the synchrotron spectrum is mostly curved. This, in addition to the steep spectrum, indicates that the radio plasma of the Kite is radiatively old. We observe that, in general, the flattest-spectrum emission is associated with the regions with the brightest X-ray emission and we estimated Spearman ranks between $\alpha$ and $I_\text{X}$ of 0.5 and 0.7 for the low- and high-band, respectively. This indicates that there is a spatial correlation between the slope of the synchrotron spectrum and the density of the thermal plasma along the arcs.\\

\begin{figure*}
\centering
 \includegraphics[width=\textwidth]{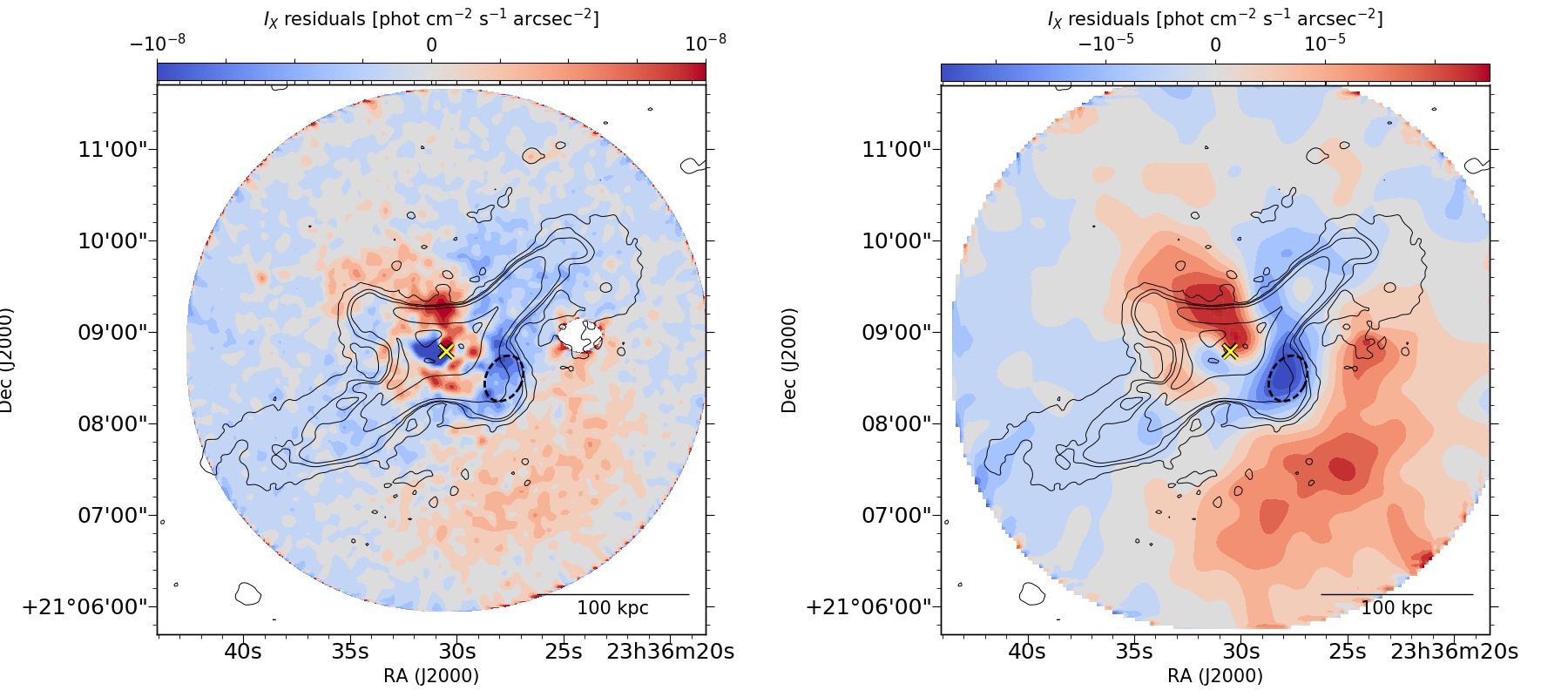}
 \caption{\label{resid}Comparison of the radio emission with the residual X-ray images produced with {\it Chandra} (left) and {\it XMM-Newton} (right) in the 0.5-2.0 keV band. Red indicates an excess with respect to the $\beta$-model, whereas blue is a deficit.  The contours are the 3, 24, 48, 192$\sigma$ of the image presented in Figure \ref{canvas}, middle panel. The images are smoothed with a gaussian filter of 2.5$''$ for {\it Chandra} and 15$''$ for {\it XMM-Newton}. the yellow cross and white mask indicate the position of, respectively, the AGN of IC5338 and the galaxy JW100. The black dashed ellipse indicate the putative position of the cavity. }
\end{figure*}
\subsection{X-ray residual maps}
Finally, we compared the radio emission with the X-ray emission similarly to the analysis performed in \citet[][]{Ignesti_2018}. The cluster is relaxed, thus the $I_\text{X}$ profile can be well described by a $\beta$-model profile \citep[e.g.,][]{Cavaliere-Fusco_1976}. Therefore, by subtracting from the observed $I_\text{X}$ a 2D $\beta$-model, fitted to the images, we can highlight substructures in the ICM, such as over-densities or cavities. We performed this analysis on both {\it Chandra} and {\it XMM-Newton} observations of the cluster using the package {\scshape sherpa} \citep[][]{Freeman_2001}. In both observations we masked the central AGN and the close spiral galaxy JW100 (see Section \ref{jw}), due to its extended X-ray emission not associated to the ICM emission \citep[][]{Poggianti_2019}. The best fit parameters are core radius $r_c$= 13.36$''$, $\beta=0.45$ for {\it Chandra} and $r_c$= 18.08$''$, $\beta=0.48$ for {\it XMM-Newton}, the results are shown in Figure \ref{resid}. We found in both the observations an excess in surface brightness (red regions in Figure \ref{resid}), which is extended from the center to the NE and crosses the northern part of the Kite, and is associated with a local excess of the thermal ICM density with respect to the mean gas density of the cluster. This structure, due to its morphology and its proximity to the cold front, has already been claimed as evidence of the sloshing of the cool core by previous works \citep[][]{Ignesti_2018,Kadam_2019}. We observe also a depression in the SW direction (dashed ellipse in Figure \ref{resid}) aligned with the jet-like features observed in the high-resolution observation at 1.4 GHz (Figure \ref{canvas}, top-right panel). The new LOFAR data show, for the first time, that the radio plasma fills this structure, thus supporting the hypothesis suggested in \citet[][]{Shin_2016} and \citet[][]{Kadam_2019} that it could be a radio-filled cavity. \citet[][]{Kadam_2019} also estimated an AGN mechanical power of 6.6$\times 10^{44}$ erg s$^{-1}$, that is in the typical range for AGN radio-mode feedback in galaxy clusters \citep[e.g.][for a review]{Gitti_2012}. Therefore, we conclude that this radio-filled cavity indicates that the radio plasma could have been originally injected by the central AGN.
\subsection{The jellyfish galaxy JW100}
\label{jw}
\begin{figure}
\centering
 \includegraphics[width=.5\textwidth]{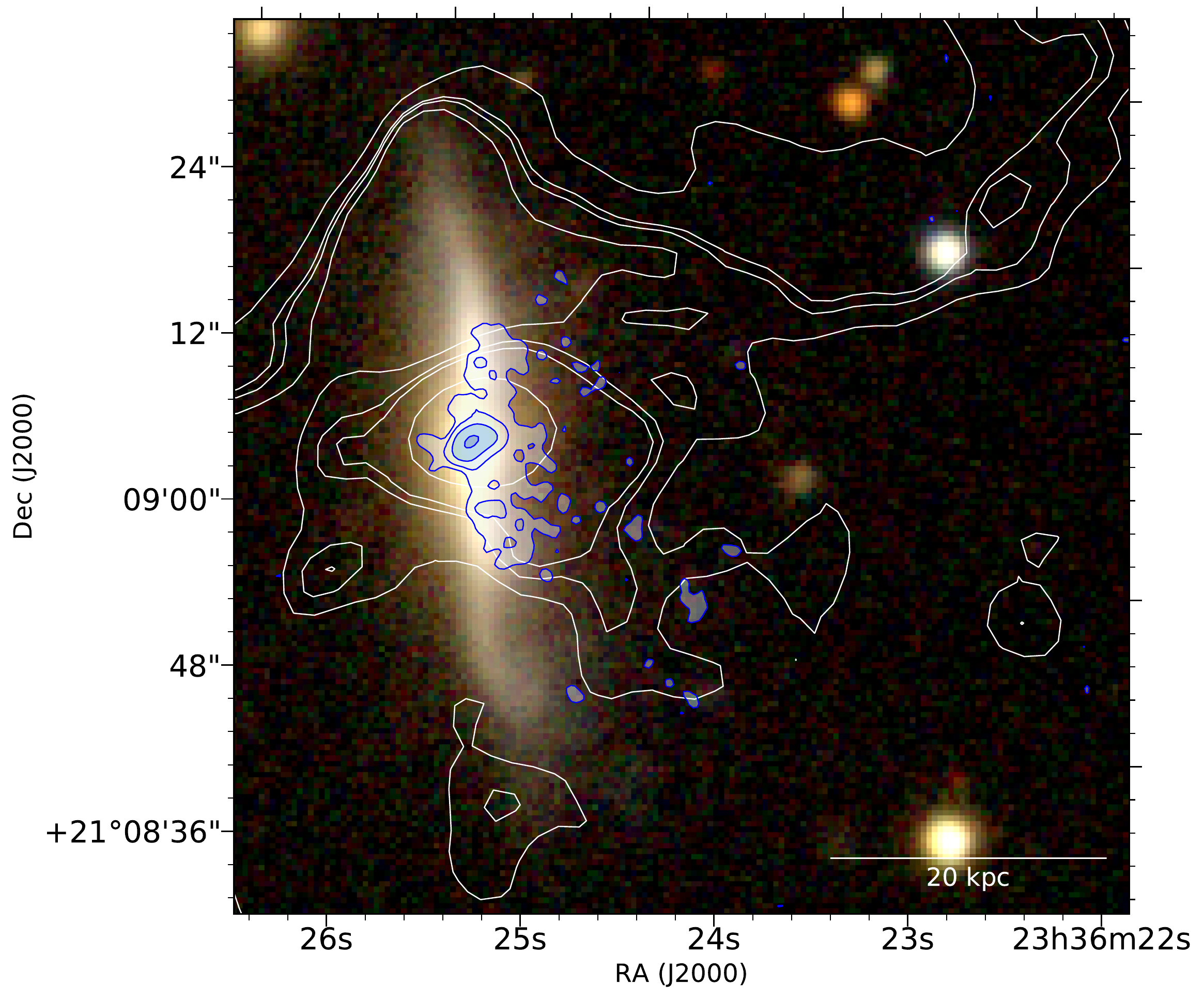}
 \caption{\label{jw100} Composite SDSS image of JW100 from bands \textit{i}, \textit{r} and \textit{g}. The blue-filled contours are 3, 6, 12, 24 $\sigma$ levels of the 1.4 GHz image, the white continuous contours are the 3, 5, 6, 12 $\sigma$ levels of the 144 MHz image (Figure \ref{canvas}).}
\end{figure}

In this paper, we also report the detection of extended radio emission associated with another galaxy of the A2626, specifically with the jellyfish galaxy JW100 \citep[also known as IC5337, $z=0.06$, ][]{Poggianti_2019}, at 144 MHz (Figure \ref{jw100}). Jellyfish galaxies are the most extreme examples of galaxies undergoing ram pressure stripping by the ICM and they represent unique laboratories to study the interplay between different gas phases and star formation. The resolution of our data allowed us to disentangle the AGN and an extended component that coincides spatially with the cold, star-forming structures observed both at 1.4 GHz and in H$\alpha$ \citep[][]{Gitti_2013b, Poggianti_2019}. We measured a total flux density of 13.3$\pm2.4$ mJy, with 4.8$\pm$1.0 mJy associated with the AGN. By subtracting the AGN contribution, we estimated that the emission from the galaxy is 8.5$\pm$2.6 mJy. In order to to estimate the spectral index, we compared these values with the flux densities measured in the same regions in the 1.4 GHz map that we used for the spectral index analysis (Section \ref{anl}). We found that both the extended emission and the AGN have a spectral index $\alpha=-0.6\pm0.1$. The k-corrected, monochromatic radio power at 144 MHz of the extended emission is 7.7$\times10^{22}$ W Hz$^{-1}$. The radio power at 1.4 GHz of the extended component is consistent with the luminosity that is expected based on the star-formation rate estimated from the H$\alpha$ emission \citep[][]{Poggianti_2019}. Hence the observed spectral index suggests that, at lower frequencies, we are observing the oldest/weakest electrons produced by the same process. Interestingly, we also detected radio emission at 144 MHz without a counterpart at 1.4 GHz. We speculate that this component could be produced by the oldest electrons injected by supernovae associated to a star formation phase prior to the one observed at 1.4 GHz. Finally, we observed that, apparently, the galaxy is linked to the northern plume of the Kite but, most likely, it is an artifact given by the resolution of our image.

\section{Discussion}
\label{disc}
\subsection{Morphology and spectral index of the Kite}
The new LOFAR data show that the diffuse radio emission is more extended than what we observe at higher frequencies (Figure \ref{canvas}, \ref{comb}). LOFAR reveals two diffuse plumes at the ends of the arcs in the NW-SE direction. In the new images, we observe that the radio surface brightness outside the arcs drops by a factor $\sim$60 within $\sim$10 kpc, whereas in the plumes the emission declines gradually. This indicates that in the arcs the non-thermal components (i.e. the magnetic field and relativistic electrons), which are traced by the radio emission, are confined. The new observation also reveals that the cold front, which appears to delimit the high-frequency radio emission \citep[][]{Ignesti_2018}, is, instead, enveloped by the low-energy, radio-emitting plasma which fills the cavity located in front of the south-west jet observed at 1.4 GHz (Figure \ref{resid}) and aligned along the major axis of IC5338 (Figure \ref{comb}, right panel). But above all, the new observation shows that the S+W and N+E arcs compose two main structures which are almost symmetric with respect to the major axis. The spectral index maps (Figure \ref{canvas_spectral}) shows that the Kite is a steep-spectrum source ($\alpha<-1.5$), in agreement  with the previous results \citep[][]{Kale_2017,Ignesti_2017}, but we observe a spatial gradient along the arcs N and S, where the spectral index steepens from their centers to their ends. Further insights into the properties of the Kite come from the residual maps (Figure \ref{resid}) and the point-to-point analysis (Figure \ref{ptp}). The spectral index map (Figure \ref{canvas_spectral}) shows that the flattest regions of the arcs are located in the north and the south with respect to the central galaxy, while the residual X-ray maps (Figure \ref{resid}) show that these regions coincide with an over-density of the thermal plasma, likely produced by the sloshing of the cool core. These two pieces of information are combined in the point-to-point analysis, where we observe that the flattest-spectrum emission comes from the regions with the brightest X-ray emission, i.e. the higher-density sloshing regions. The point-to-point analysis also reveals significant curvature in the spectrum, which is indicative of radiatively old plasma. \\
\subsection{A new scenario for the origin of the Kite}
\begin{figure*}
\begin{multicols}{2}
\includegraphics[width=.5\textwidth]{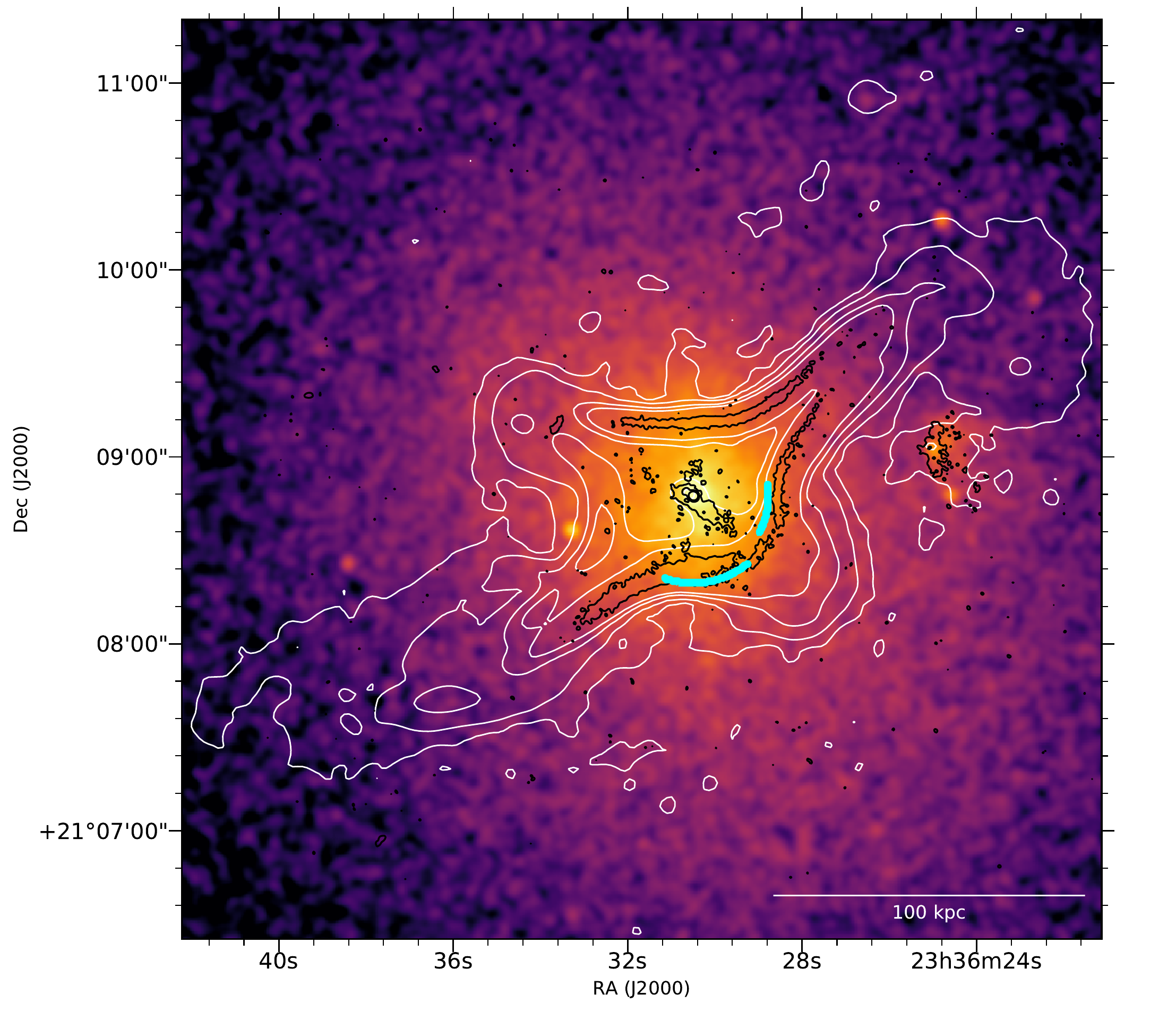}\par
\includegraphics[width=.5\textwidth]{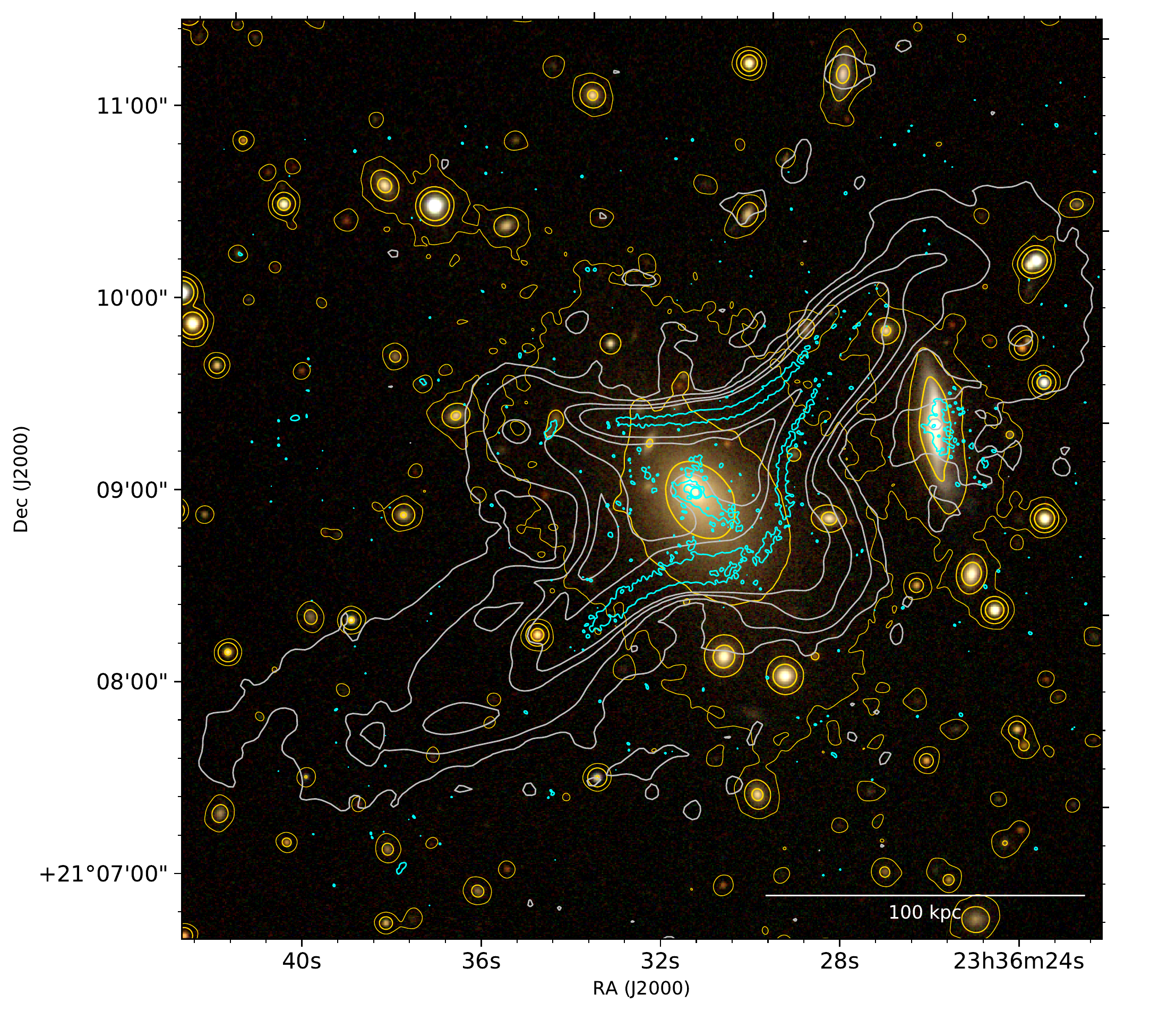}\par
 \end{multicols}
 \caption{\label{comb}Left: {\it Chandra} X-ray image of A2626 in the 0.5-2.0 keV band with the 3, 24, 96, 200$\sigma$ level contours of the emission observed at 144 MHz in white ( Figure \ref{canvas}, middle panel) and the 3, 24, 96$\sigma$ contours of the emission at 1.4 GHz in black (Figure \ref{canvas}, top-right panel). In cyan is reported the position of the cold front observed in \citet[][]{Ignesti_2018}; Right: Composite SDSS image from bands \textit{i}, \textit{r} and \textit{g} with the 3, 24, 96, 200$\sigma$ contours of the emission at 144 MHz in silver (Figure \ref{canvas}, middle panel), the 3, 24, 96$\sigma$ contours of the emission at 1.4 GHz in cyan (Figure \ref{canvas}, top-right panel) and SDSS composite image smoothed with a Gaussian kernel of standard deviation 1 arcsecond at 6, 30, 150 times the RMS noise in the smoothed image (0.015 maggies) in yellow. }
 \end{figure*}

On the basis of our results, we speculate on the origin of the Kite. On the one hand, we claim that is very unlikely the Kite is a mini-halo. Moreover, the morphology revealed by LOFAR also discourages the lensing hypothesis \citep[previously invoked in ][]{Kale_2017} because produce such a blend of collimated arcs and large plumes would require implausible combinations of lenses and scattering screens. On the other hand, in the new image at 144 MHz the Kite shows remarkable similarities with two specific classes of radio galaxies, the Z-shaped \citep[e.g.,][]{Zier_2005,Hardcastle_2019} and X-shaped radio galaxies \citep[e.g.,][]{Lal_2007,Bruno_2019,Cotton_2020}. The Kite shares the double-axis morphology with these radio sources, where the wings (i.e. the $``$plumes" of the Kite, see Figure \ref{scheme}) are more extended than the lobes.
Moreover, the central, giant elliptical galaxy IC5338, is elongated in the same direction of the putative minor axis of the Kite from NW to SW. The extent of this is shown by the smoothed SDSS contours in the right panel of Figure \ref{comb} (right panel), where a Gaussian kernel of width 1 arcsecond was used to improve sensitivity to the extended stellar emission and avoid confusion with surrounding point sources. Furthermore, the radio filled cavity (Figures \ref{resid}) and the radio jets originating from the central AGN (Figure \ref{comb}) are aligned along the same NW to SE axis. Therefore, we suggest that originally the Kite was a radio galaxy whose central AGN produced the S-W cavity. 
In this framework, the peculiar, double-axis morphology of the Kite can be then explained by the backflow model \citep[e.g.,][]{Leahy_1984, Hodges_2011}. This hydrodynamical model, which is one of the possibilities that have been invoked to explain X-shaped radio galaxies, suggests that the plasma injected by the AGN along the jets can be redirected by the hot-spots or high-pressure environment back to the galaxy. In the case of axisymmetric backflow, \citet[][]{Leahy_1984} noted that this process can cause the emission to grow in directions perpendicular to the angle at which the plasma impacts with the hot inter-stellar medium (ISM) of the galaxy. Therefore, for the case of the Kite, we suggest that the plasma was injected in the NE-SW direction, along the projected major axis of the IC5338 (Figure \ref{comb}, right panel), to subsequently flow back and be redirected in the NW-SE directions to finally form the plumes. In this scenario, the hot, X-ray emitting ISM should follow the same elliptical geometry of the optical emission. However, this is not immediately observed in the Chandra image (Figure \ref{comb}, left panel), possibly due to the complex dynamics of the ISM that could be induced by the AGN outburst.\\

However, there are several properties of the Kite that make this object peculiar and different from standard radio galaxies. First, the overall spectrum of the Kite ($\alpha<-1.5$) is steeper than the typical spectral index of active X-shaped radio galaxies \citep[$\alpha>-1.5$, e.g.,][]{Lal_2007}, which indicates that the plasma is radiatively-old. The steep spectrum measured in the 144-610 MHz band (Figure \ref{ptp}) suggests that the break frequency of the spectrum, $\nu_\text{br}$, is below 144 MHz and, hence, that we are observing only the radiatively-old part of the of the synchrotron spectrum. Therefore, by assuming a putative $\nu_\text{br}<100$ MHz and under the hypothesis that the local magnetic field, $B$, is in the range 5-15 $\mu$G, which are typical for radio galaxies \citep[e.g.,][]{Croston_2005}, we can derive a constraint for the radiative age of the arcs, $t_\text{rad}$ as:
\begin{equation}
 t_\text{rad}\simeq 3.2\times10^{10}\frac{B^{1/2}}{B^2+B_\text{CMB}^2}\left[(1+z)\nu_\text{br}\right]^{-1/2}\text{ yr}
\end{equation}
where $B$ and the equivalent magnetic field of the cosmic microwave background (CMB) $B_\text{CMB}=3.25(1+z)^2=3.6$ are expressed in units of $\mu$G and $\nu_\text{br}$ is expressed in units of MHz. Under these assumptions, we determine the lower limit to the radiative age of the plasma to be between $5\times10^7$ and $2\times10^8$ yrs, which confirms that we are observing radio emission from old, fossil plasma. This time-scale is not far from the typical duty-cycle of an AGN \citep[$\sim10^8$ years, e.g.,][]{Morganti_2017} and, thus, we suggest that the active phase that injected the relativistic plasma of the Kite likely ended not too long ago.\\

Old radio plasma sitting in the center of clusters is expected to be uplifted due to buoyancy in a certain timescale. This provides complementary constraints to the age of the plasma in the Kite. The facts that the source preserves its X-shaped morphology and the plumes still sit at $\sim100$ kpc from the central AGN implies that the plasma did not have enough time to evolve buoyantly. According to \citet[][]{Churazov_2001}, the buoyant velocity of a bubble can be estimated as:
\begin{equation} 
v_b \simeq v_k\sqrt{\frac{2V}{RSC}}
\label{bu}
\end{equation}
where $v_k$ is the Keplerian velocity, which we approximate to the cluster dispersion velocity $\sigma_v \simeq 680$ km s$^{-1}$ \citep[][]{Cava_2009}, $S$ is the cross-section of the bubble, $V$ its volume, $R$ the distance from the central AGN and $C\sim 0.6$ is the drag coefficient. As a reference, by assuming the plumes as spherical bubbles with $r=25$ kpc at a distance $R\simeq100$ kpc from the AGN, Equation \ref{bu} gives us a reference buoyant velocity $v_b \sim 700$ km s$^{-1}$. The mean ICM temperature at 100 kpc from the AGN is 3.4 keV \citep[][]{Ignesti_2018} with a corresponding sound speed of $\sim940$ km s$^{-1}$, hence the bubble would be rise sub-sonically in the ICM. This bouyant velocity entails a upper limit to the dynamical time-scale of the plumes $<1.5\times 10^8\left[\text{sin}(\theta)\right]^{-1.5}$ yr, where $\theta$ is the angle between plumes and the line of sight, which is in line with previous constraints and suggests that the plasma is not much older than a typical AGN duty cycle.\\

Secondly, contrary to what is generally observed or expected in back-flowing radio galaxies, in the Kite the plumes have a projected length in the NW-SE direction that is almost double of the length of the jets, which we assumed originally directed in the NE-SW direction. This could be both due to local environmental properties (i.e. a stronger environmental pressure that restricted the expansion of the plasma in the jet direction) combined with bouyant motions of the radio plasma in the plume or complex projection effects due to a possible inclination of the Kite with respect to the plane of the sky. The latter hypothesis is partially supported by the spread of the color-color diagram (Figure \ref{ptp}), which could be due to the mixing of plasma with different radiative ages, hence spectral indices, along the line of sight.\\

Finally, the most obvious difference with X-shaped galaxies is that the arcs of the Kite are clearly detached from the central AGN (Figure \ref{canvas}, bottom-left panel) and, on the contrary to what is observed for X-shaped galaxies, the brightest part of the Kite are the arcs, where we observe also a gradient in spectral index and a spatial correlation between the $\alpha$ and the X-ray emission (Figure \ref{ptp}).
\subsection{The role of the ICM dynamics}
We interpreted the presence of edge-brightened features as indication of a role played by the ICM dynamics in the origin of this radio source. Interestingly, the radio spectral properties of the Kite resemble those of the radio phoenix sources, an exotic class of diffuse radio sources produced by the interplay of fossil radio plasma with shocks and/or ICM motions \citep[e.g.,][]{Clarke_2013, Mandal_2020}. Specifically, the arcs of the Kite resemble the complex, filamentary radio morphology of the radio phoenices, whose filaments have patchy distributions of increased surface brightness, and synchrotron emission typically that follow a curved spectra. To tie together these scenarios, we speculate that after the end of the AGN activity, when the plumes were already formed, the sloshing of gas in the cool core has interacted with the fossil radio plasma (Figure \ref{resid}, \ref{comb} left panel). The typical sloshing crossing-time is of the order of 10$^8$ years \citep[e.g.,][]{Ascasibar_2006}. Therefore, according to our estimates, the Kite would be old enough to allow the sloshing to impact significantly on its radio emission. On the basis of the temperature maps presented in literature \citep[][]{Ignesti_2018,Lagana_2019,Kadam_2019}, we suggest that the ICM motion could have interacted with the fossil plasma mostly in the N-S direction. The motion of the thermal plasma could have enhanced the radio emissivity at the edges of the radio galaxy by compressing the relativistic plasma and producing the bright arcs of the Kite.
The compression of radio plasma can efficiently enhance its radio emission by a factor $\delta I$ of:
\begin{equation}
 \delta I \propto x^{-\frac{2}{3}\delta+1}
\end{equation}
where $x$ is the compression factor and $\delta=2\alpha-1$ is the index of the electron spectrum $N_{e}(E)\propto E^{\delta}$  \citep[e.g.,][]{Markevitch_2005}. In the case of the Kite, by using a power-law spectrum instead of a curved spectrum, we can derive a first-order estimate of the effects of compression. For $\alpha<-1.5$, hence $\delta<-4$, even a small compression $x\simeq 1.2$ can results in a significant enhancement $\delta I>$2. At the cold front, where the compression would be more evident, \citet[][]{Kadam_2019} measured values of $x$ of $1.57\pm0.08$ and $2.06\pm0.44$, that would correspond to a $\delta I\simeq6$.\\

The local increase of the magnetic field due to the compression impacts also on the spectrum of the emission by shifting the critical frequency of emission of the electrons at higher frequencies. For a curved synchrotron spectrum, as we observe here for the Kite, this compression, by moving the break frequency to higher values, can result in a flattening of the observed spectrum. This effect should be more noticeable where the ICM compression, and consequently also $I_\text{X}$, is stronger. This is in agreement with the point-to-point analysis (Figure \ref{ptp}), which shows that the flattening of the spectrum is most evident in the regions with the highest $I_\text{X}$, and, thus, it supports a scenario where the radio emission is driven by the local properties of the thermal ICM.\\
\subsection{The role of the central AGN}
The steep spectral index of the arcs and the color-color plot indicates that the arcs were formed after the end of the duty cycle, when the central AGN entered a low-activity state. Indeed IC5338 does not show properties of AGN emission from the optical (SDSS) or infrared (WISE) photometry. Furthermore, a recent catalogue identifying quasars from photometric data in SDSS and WISE indicate that galaxies in A2626 have a very low probability of being quasars \citep[][]{Clarke_2019}. However, an AGN coincident with IC5338 is clearly detected at 144 MHz (Figure \ref{scheme}) . We note at the centre of the galaxy there are two compact cores shown in the SDSS image. The core to the SW is coincident with the 1.4 GHz radio emission (shown by cyan contours in the right image of Figure \ref{comb}) and it exhibits hard X-ray emission \citep[][]{Wong_2008}, thus it is likely the host of the AGN. The presence of these two cores in close proximity suggests a merger event had occurred to form IC5338, and there may be significant ongoing tidal forces between the two cores. Evidence for a merger event helps to provide a mechanisms for the supply and accretion of gas onto the AGN in the SW core \citep[e.g.,][for a recent result]{Ellison_2019}.\\

The spectral index of the AGN ($\alpha=$-0.8$\pm0.1$, Figure \ref{canvas_spectral}) indicates that the supermassive black hole is currently accreting material. Furthermore, at higher frequencies the VLA high-resolution observations presented by \citet[][]{Gitti_2013b} (Figure \ref{canvas}, \ref{comb}) show the existence of a small ($\sim 8$ kpc), jet-like feature leaving the nucleus of the central galaxy, and, thus, confirming that the AGN has entered a new activity phase. The non-detection of an AGN in optical or infrared data hence suggests that it may be obscured by the accretion disk and potentially intra-cluster dust along the line of sight or it may be accreting via a radiatively-inefficient mode. Therefore, we can not exclude that the starting of a new duty cycle by the central AGN could have played a role in the origin of the Kite by lunching shocks in the old radio plasma and/or by further compressing the fossil plasma from the inside in the NE-SW direction.

\section{Summary and Conclusions}
We presented the analysis of the a LOFAR observation at 144 MHz of the galaxy cluster Abell 2626 and its puzzling central radio source, also known as the Kite. The new LOFAR data have radically changed the general picture of the Kite. We detected new components of the radio emission in the form of plumes located at the ends of the arcs, which almost double the total length of the radio source to $\sim220$ kpc with respect to the higher frequencies images. We combined the new data with the archival datasets at 1.4 GHz and 610 MHz to map the spectral index of the Kite from 144 MHz to 1.4 GHz. We find that the spectral index of the arcs steepens at higher frequencies from $\alpha\simeq-1.5$ to $\alpha<-2$ with a steepening trend from their centers to their ends. Then we compared the radio and the X-ray emission and we found that the radio plasma fills a putative cavity located beyond the S-W junction of the arcs which is aligned along the major axis of IC5338. Finally, by performing a point-to-point analysis we found a spatial correlation between the spectral index and the X-ray surface brightness, which is a proxy of the local ICM thermal plasma density, where the flattest-spectrum parts of the arcs coincide with the brightest X-ray regions. This result highlights that the properties of the radio emission of the Kite are driven by the local properties of the thermal ICM. On the basis of our findings, we propose that the Kite could be a relic radio galaxy whose lobes have been shaped by the motion of the gas sloshing in the cool core. Specifically,  the morphology observed for the first time by LOFAR and the steep spectrum of the lobes ($\alpha<-2$) suggest that the Kite was originally an X-shaped radio galaxy, likely produced by the back-flow of the radio plasma injected by the AGN that was redirected by the galactic hot ISM along the plumes. The spectral index and the size of the lobes set the lower limit for age of the system between $5\times10^7$ and $2\times10^8$ years. The curved radio spectrum and the spatial connection with the cold front suggest that, subsequently after the end of the AGN activity, the fossil radio plasma has been compressed by the ICM motion that enhanced locally the emissivity of the radio arcs and flattened their spectral index. Such a combination of phenomena can also explain its uniqueness among the radio sources because only a very favorable combination of different mechanisms could have resulted in such a spectacular radio source. This scenario can be now tested with tailored numerical simulations, which can explore the physical boundaries of the back-flow and the ICM compression, and future, high-resolution observations at 144 MHz carried out with the LOFAR international stations, which could provide us images of the Kite with a resolution of 0.3$''$. 
\section*{Acknowledgments}
We thank the Referee for their suggestions that improved the presentation of this work. AI thanks L. Bruno for useful discussions. This research made use of Astropy, a community-developed core Python package for Astronomy \citep[][]{astropy_2013, astropy_2018}, and APLpy, an open-source plotting package for Python \citep[][]{Robitaille_2012}. GB and RC acknowledge support from INAF through mainstream program "Galaxy clusters science with LOFAR". MJH acknowledges support from the UK Science and Technology Facilities Council (ST/R000905/1). RJvW and AB acknowledges support from the VIDI research programme with project number 639.042.729, which is financed by the Netherlands Organisation for Scientific Research (NWO). AOC gratefully acknowledge support from the UK Research $\&$ Innovation Science $\&$ Technology Facilities Council (UKRI-STFC). GDG acknowledges support from the ERC Starting Grant ClusterWeb 804208. VC acknowledges support from the Alexander von Humboldt Foundation. AMS is supported by the UK's Alan Turing Institute under grant 2TAFFP/100012.
LOFAR \citep[][]{vanHaarlem_2013}  is the Low Frequency Array designed and constructed by
ASTRON. It has observing, data processing, and data storage facilities in several countries,
which are owned by various parties (each with their own funding sources), and that are
collectively operated by the ILT foundation under a joint scientific policy. The ILT resources
have benefited from the following recent major funding sources: CNRS-INSU, Observatoire de
Paris and Université d'Orléans, France; BMBF, MIWF-NRW, MPG, Germany; Science
Foundation Ireland (SFI), Department of Business, Enterprise and Innovation (DBEI), Ireland;
NWO, The Netherlands; The Science and Technology Facilities Council, UK; Ministry of
Science and Higher Education, Poland; The Istituto Nazionale di Astrofisica (INAF), Italy.
This research made use of the Dutch national e-infrastructure with support of the SURF
Cooperative (e-infra 180169) and the LOFAR e-infra group. The Jülich LOFAR Long Term
Archive and the German LOFAR network are both coordinated and operated by the Jülich
Supercomputing Centre (JSC), and computing resources on the supercomputer JUWELS at JSC
were provided by the Gauss Centre for Supercomputing e.V. (grant CHTB00) through the John
von Neumann Institute for Computing (NIC).
This research made use of the University of Hertfordshire high-performance computing facility
and the LOFAR-UK computing facility located at the University of Hertfordshire and supported
by STFC [ST/P000096/1], and of the Italian LOFAR IT computing infrastructure supported and
operated by INAF, and by the Physics Department of Turin university (under an agreement with
Consorzio Interuniversitario per la Fisica Spaziale) at the C3S Supercomputing Centre, Italy. We thank the staff of the GMRT that made these observations possible. GMRT is run by the National Centre for Radio Astrophysics of the Tata Institute of Fundamental Research.

\bibliographystyle{aa}
\bibliography{bibliography.bib}
\end{document}